# On-Chip Quantum Interference from a Single Silicon Ring Resonator Source

Stefan F. Preble,[1] Michael L. Fanto,[2] Jeffrey A. Steidle,[1] Christopher C. Tison,[2] Gregory A. Howland,[2] Zihao Wang,[1] and Paul M. Alsing[2]

[1]Rochester Institute of Technology, Microsystems Engineering, 168 Lomb Memorial Dr. Rochester, NY 14623
[2]Air Force Research Laboratory, 525 Brooks Rd., Rome, NY 13441

Here we demonstrate quantum interference of photons on a Silicon chip produced from a single ring resonator photon source. The source is seamlessly integrated with a Mach-Zehnder interferometer, which path entangles degenerate bi-photons produced via spontaneous four wave mixing in the Silicon ring resonator. The resulting bi-photon N00N state is controlled by varying the relative phase of the integrated Mach-Zehnder interferometer, resulting in high two-photon interference visibilities of V~96%. Furthermore, we show that the interference can be produced using pump wavelengths tuned to all of the ring resonances accessible with our tunable lasers (C+L band). This work is a key demonstration towards the simplified integration of multiple photon sources and quantum circuits together on a monolithic chip, in turn, enabling quantum information chips with much greater complexity and functionality.

Silicon Photonics is proving to be a promising platform for quantum information processing applications [1–5]. Photon sources with high brightness and spectral purity, low noise and compact footprints have been demonstrated using spontaneous four wave mixing (SFWM) in ring or microdisk resonators [6–12]. These sources have been used to demonstrate time-energy entanglement in off-chip setups and photonic integrated circuits [13,14]. High efficiency superconducting nanowire single photon detectors have also been integrated with waveguide based quantum circuits [4,15], enabling the integration of all of the key components of a quantum information processor. However, complex quantum information processors will require many sources to be integrated and entangled together in waveguide based circuits. Recently the first steps have been taken towards this with the integration of photon sources and entanglement circuits on a single Silicon chip [16,17]. However, these demonstrations have required multiple, independent, photon sources to realize two photon entanglement. Ergo, the inherent fabrication variations in the sources will yield photon pairs that are potentially distinguishable. This has been alleviated partially in the case of resonant photon sources by using a pulsed pump laser but it is advantageous to be able to make use of narrow linewidth continuous wave pump lasers in order to ensure long coherence lengths of the entangled photon states.

Here we demonstrate the entanglement of photon pairs produced from a *single* ring resonator photon source. This ensures that the entangled photons will be indistinguishable since they are generated within the same device. Furthermore, since entanglement is realized from a single device the number of photon source devices in an entangling circuit can be halved. Furthermore, we will show here that the resonant photon source can be operated with pump laser wavelengths that span ~80nm, in turn, enabling wavelength multiplexing in future systems. Consequently, this work is a key advance in the simplification of integrated quantum information systems and will enable complex integration of sources and entanglement circuits together for a wide variety of quantum technologies, such as, quantum key distribution, and possibly even quantum computation.

In order to use a single ring resonator as an entangled photon source, we excite the resonator bi-directionally. This is seen in Fig. 1(a) where two pump lasers [Pump 1 (blue) and Pump 2 (red) – each tuned to a resonance of the ring resonator] are launched into the Silicon chip. On the chip, they are immediately split with a y-splitter [18] and routed into a loop. The counter-propagating pumps are then coupled into a ring resonator photon source [Radius $R = 18.5\mu$m, waveguide dimensions of $W = 500$nm by $H = 220$nm, and a waveguide to ring gap of $g = 150$nm] where they induce SFWM in opposite directions, which generates entangled photon pairs. The ring resonator was designed to have zero dispersion, therefore, when the pump frequencies are tuned to resonances that are equally spaced from a central resonance, degenerate bi-photons are produced as dictated by the energy conservation relation: $E_{\text{Pump}_1} + E_{\text{Pump}_2} = E_{\text{Bi-Photon}}$ (depicted by the virtual energy level diagram in Fig.1(a)). We have recently shown that this photon generation process results in bright, low noise, degenerate bi-photons [10].

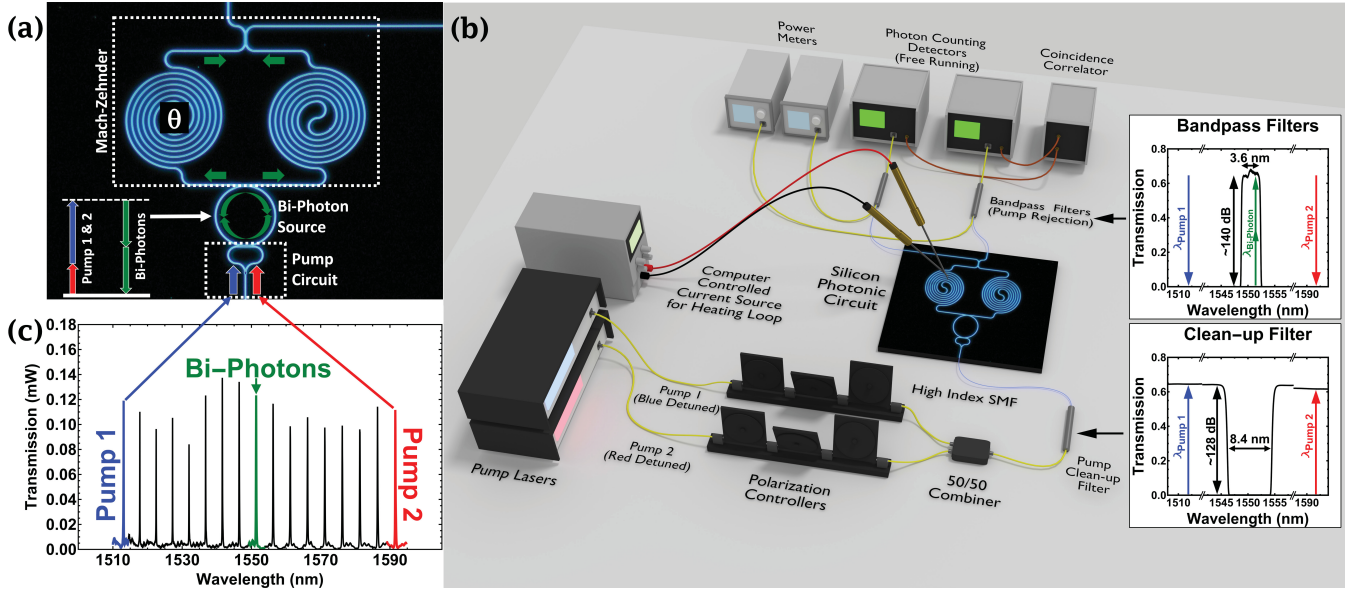

FIG. 1. (a) Dark-field microscope image of the quantum circuit on the Silicon-On-Insulator chip consisting of an integrated pump splitting circuit, ring resonator (Q~15k, FSR~5nm) entangled bi-photon source and Mach-Zehnder analysis circuit. (b) Schematic of the complete experimental setup. A pair of tunable lasers, Pump 1 (Blue Detuned - New Focus 6328) and Pump 2 (Red Detuned - Agilent 81642A) with 1mW of power, are polarized, combined and passed through pump clean-up filters (Clean-up Filter inset: ~128 dB rejection). A short section of High Index fiber (Nufern UHNA-7) was fusion spliced to optical fiber (SMF-28) to efficiently mode match to the Silicon waveguide achieving a total insertion loss of ~5dB (we have demonstrated <1.25dB fiber-chip coupling loss using this method before [10]). The relative phase (θ) in the Mach-Zehnder was controlled by heating one of the spirals (by varying the current through shorted needle probes in contact with the spiral). Photons from the two outputs passed through bandpass pump rejection filters (Bandpass Filters inset: ~140 dB). The rejected pumps are sent to optical power meters to measure the classical signatures. The bi-photons are detected by free-running InGaAs Avalanche Photodiodes (idQuantique ID210 – 10% Efficiency, ID230 – 25% Efficiency) and correlated using a time-to-digital converter (Picoharp 300). (c) Transmission spectrum of the ring resonator (as measured from one of the outputs of the Mach-Zehnder interferometer when the relative phase is tuned for ~50/50 splitting) indicating the wavelengths of the two pump lasers [Pump 1 (blue) and Pump 2 (red)] as well as the resonant wavelength of the resulting bi-photons ("green").

For this experiment, we have set the generation rate to be sufficiently low so that higher order generation events are unlikely to occur. In such a regime, we can approximate the state generated within the ring resonator to be:

$$|\psi\rangle_{ring} = \frac{1}{2}\left[\hat{a}_{cw}^{\dagger\,2} + \hat{a}_{ccw}^{\dagger\,2}\right]|vac\rangle \quad (1)$$

where $\hat{a}_{cw}^{\dagger}$ and $\hat{a}_{ccw}^{\dagger}$ are independent creation operators for clockwise and counter-clockwise propagating bi-photons, respectively. This unique pumping configuration results in a two-photon N00N state (N=2), which is a well-known form of entangled state that exhibits a factor of $N$ sensitivity to relative phase changes over classical light.

This state is coupled out of the resonator, and into a Mach-Zehnder analysis circuit for confirmation (Fig. 1(a)). The circuit is composed of two legs of waveguide which are spiraled to increase their total path length (~1mm). One of the spirals is thermally tuned to induce a relative phase shift [θ] between the two paths. Combining the legs of the Mach-Zehnder onto a 50/50 directional coupler completes the analysis resulting in the phase dependent output state:

$$|\psi\rangle_{out} = \frac{1}{4}e^{-i\theta}\left[\begin{array}{c}\hat{a}_{cw}^{\dagger\,2} - \hat{a}_{ccw}^{\dagger\,2} + 2i\hat{a}_{cw}^{\dagger}\hat{a}_{ccw}^{\dagger} + \\ \left(-\hat{a}_{cw}^{\dagger\,2} + \hat{a}_{ccw}^{\dagger\,2} + 2i\hat{a}_{cw}^{\dagger}\hat{a}_{ccw}^{\dagger}\right)e^{2i\theta}\end{array}\right]|vac\rangle \quad (2)$$

This state oscillates between high coincidence and zero coincidence as a function of phase delay as follows:

$$|\psi\rangle_{coinc.} = i\hat{a}_{cw}^{\dagger}\hat{a}_{ccw}^{\dagger}|vac\rangle, \quad \theta = \pi n, \quad n = 0,1,2,\ldots \quad (3)$$

$$|\psi\rangle_{no\,coinc.} = \frac{i}{2}\left[\hat{a}_{cw}^{\dagger\,2} - \hat{a}_{ccw}^{\dagger\,2}\right]|vac\rangle, \quad \theta = \pi\left(n+\frac{1}{2}\right) \quad (4)$$
$$n = 0,1,2,\ldots$$

The distinguishing feature of this two-photon interference is that it oscillates at twice the frequency as classical light. Specifically, the bi-photons interfere at the output of the Mach-Zehnder so that they either bunch together (no coincidences) or anti-bunch (coincidences) at phase shifts that are multiples of π, whereas classical light in the Mach-Zehnder experiences constructive/destructive interference over multiples of 2π. Consequently, the observation of bi-photon coincidences that oscillate at twice the rate as is seen with a classical signal is a clear indication of bi-photon N00N state generation and interference in the device.

This is seen in Fig. 2 where the measured classical signal and coincidences are plotted. The classical light output from the two ports of the Mach-Zehnder is shown in Fig. 2(a). As the relative phase is changed, the signals cycle between destructive and constructive interference and it is clear that the coincidences shown in Fig. 2(b) oscillate at twice the frequency, supporting our theorized output state in Eqn. (2). Consequently, counter-propagating entangled bi-photons must have been generated in the ring resonator in order to observe the high visibility coincidence oscillations observed here.

In order to show that the two-photon interference is a direct result of the generation of an entangled bi-photon quantum state in a single ring resonator, we shifted the wavelength of the "Pump 2" laser by one resonant peak towards the central bi-photon resonance (seen in Fig. 1(c)). In this case, energy conservation dictates that the generated bi-photons will no longer be generated at the symmetrically centered resonance ($\lambda_{Bi-photon} \sim 1551nm$). Instead any bi-photons that could be generated would have to be at a wavelength that does not correspond to an actual resonance, which strongly inhibits their generation [11]. We do note though that if any bi-photons were to be generated that they would still fall within the transmission window given by the bandpass pump rejection filters (Fig. 1(b)) and as a result they would be measured. However, as is seen in Fig. 2(b) (Incoherent-"orange-line"), no coincidences are observed and it can be concluded that entangled bi-photons were not generated even though both pump lasers are still producing photons. Specifically, the only photons that are generated are from the independent four wave mixing actions of each pump laser. These independently generated photon pairs must be spectrally non-degenerate and as a result, when they are only measured within the spectral window of the pump rejection filter, the photons are effectively incoherent (e.g. thermal); consequently, they do not interfere with each other.

The observed photon interference in Fig. 2(b) has a measured visibility of $V_{(raw)}= 93.3 \pm 2.0\%$. When the accidentals are subtracted (which is common practice and is done to remove the noise from the detectors and other photon generation processes [17]) the visibility is observed to be $V=96.0 \pm 2.1\%$. These high visibilities indicate that the observed coincidences arise from the bi-photon entanglement discussed in this letter. We do note though that the visibilities could be improved upon. First, it is seen that the classically measured visibilities in Fig. 2(a) are lower than desired. This is due to un-optimized directional couplers (e.g. length and waveguide separation gap) in the Mach-Zehnder interferometer. Another source of noise is seen in the coincidence data around a phase shift of $\sim\pi$. This noise is due to the thermal/mechanical stability of the setup; particularly, the electrical needle probes used to locally heat one leg of the Mach-Zehnder interferometer. Since the data was measured over many hours, the probes inevitably shifted position, which strongly affects the coincidences at this point because they occur over the high sensitivity quadrature points of the classical signal.

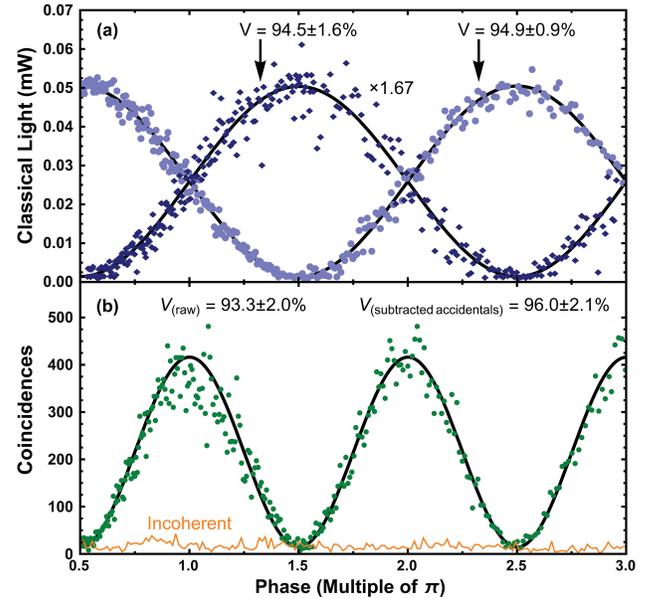

FIG. 2. (a) Classical laser light from the two outputs of the silicon circuit (blue diamonds and purple circles) as the relative phase of the Mach-Zehnder was varied. The heating power was converted to phase using a fit of the applied heating power and assuming the phase linearly depends on temperature. The blue-diamond data was scaled by a factor of x1.67 for ease of comparison. This path experienced a higher loss on the chip because the waveguide was physically longer. This longer waveguide was required in order to couple light in/out of the chip using the three optical fibers, each controlled by an independent piezo stage. (b) Coincidences between the two paths measured for ninety-seconds (green circles). The coincidences were measured with a 32ps time resolution and integrated over 224ps in order to account for the timing jitter of the Avalanche Photodiodes. This integration range was found to maximize the observed visibility. The accidentals were obtained by averaging the time correlations far away from the coincidence peak over a time interval of 320ns. The calculated error is obtained from the standard deviation of the accidentals over the same range. (orange line – Incoherent): Coincidence counts when the pumps are tuned asymmetrically about the central resonance, resulting in independent, incoherent, photon pairs.

The peak coincidences observed were also limited by loss within the device and the detectors. The free running InGaAs photon detectors have relatively low dark count rates [~2.5kHz for the ID210, and ~200 for the ID230] but only when operated with long dead times [50μs-id210, 25μs-id230]. We characterized the detectors and for the photon flux detected we determined that approximately 50% of the photons were being lost within the dead time of the detector. Consequently, the actual coincidences could increase by approximately four times with better detectors. It should also be noted that the ring resonator photon generation device itself has an effective photon-pair loss.

Specifically, the bi-photons generated within the microring cavity may be lost through different ports of the ring resonator as recently pointed out by Vernon & Sipe in [19]. However, this is inherently true for any critically coupled resonator – even one that is just coupled to a single waveguide [19].

Noise is also produced by non-degenerate photons produced by the pumps acting independently of each other. We previously observed that a single pump will produce a comb of photon pairs [10]. Therefore, each pump spontaneously produces an incoherent photon at the bi-photon resonance. In order to minimize this effect it is key to optimize the relative powers of the two pump lasers. Specifically, if the power of one laser is stronger than the other one, more non-degenerate photon pairs will be produced by that stronger pump. Consequently, it is important to match the two pump powers to maximize the Coincidental/Accidental Ratio (CAR) of the bi-photons. For all of the measurements shown in this letter this optimization was done and resulted in a CAR of ~80 when both pumps each had 1mW of power (~350uW each at the chip). Higher CAR's can be obtained at lower pump power levels but we selected this power as it gave a reasonable tradeoff between noise and photon flux. We also note that noise is produced from Raman processes in the UHNA7 fiber and the glass cladding surrounding the Silicon waveguide [10].

While the presence of multiple resonance modes is a challenge from a noise standpoint, the multiple resonances provide flexibility in how the bi-photons are produced. Specifically, all of the resonances accessible with our tunable lasers can be used to produce bi-photon pairs, provided the two pump lasers are tuned such that energy conservation is met for the central bi-photon resonance. This requires the dispersion to be relatively flat, but [as seen in Fig. 3(a-f)] when the two pump lasers are symmetrically tuned to all resonances over a range spanning up to $\Delta\lambda=\lambda_2-\lambda_1=78.5$nm and down to $\Delta\lambda=19.4$ (which is the closest the pumps could be placed together due to the limitations of the filter bandwidths (Fig. 1(b) insets)), high visibility two photon interference is observed. We do note though that there are some variations in both the peak coincidences and the accidentals that do yield variability in the observed visibility. We believe this is due to the weak wavelength dependence observed for the entire circuit (Fig. 1(c)), which is likely a result of the dispersion of the directional couplers. Regardless, the ability to use any resonant wavelengths to produce the two photon interference enables flexibility for future wavelength division multiplexed integrated quantum circuits.

Another important consideration is how selective the wavelength dependence of the two photon interference is. We measured the spectral selectivity of the resonator by setting the relative phase of the Mach-Zehnder to $\pi$ (i.e. peak coincidences) and then scanned the wavelengths of both pump lasers, as seen in Fig. 3(g-h). We find that the coincidences only occur when the two pump wavelengths coincide with the energy conservation relation of the spontaneous four wave mixing process. Here this corresponds to $\Delta\lambda_{FWHM}\sim0.13$nm, which closely matches the inherent spectral width of the ring resonances and supports the conclusion that the bi-photon bandwidth is dictated by the resonator [11].

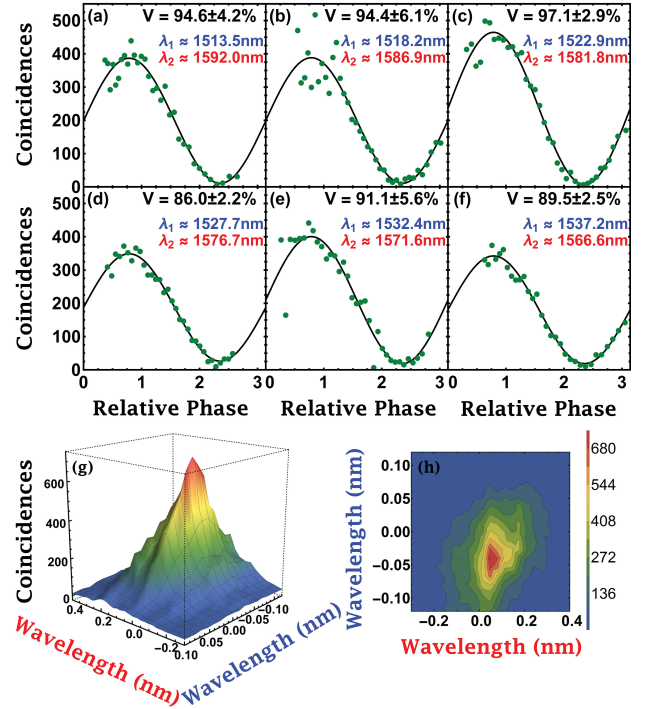

FIG. 3. (a-f) Coincidences for different pump laser wavelengths. (g-h) Dependence of the coincidences on the wavelengths of the two pump lasers. The scales are different because the New Focus laser (blue scale with range -0.1 to 0.1nm) has a very limited fine-tuning range.

In conclusion we have demonstrated quantum interference of photons produced from a single Silicon ring resonator photon source. This was achieved by exploiting the inherent degeneracy of the clockwise and counterclockwise modes of a ring resonator. This effectively doubles the functionality of a single ring resonator photon source and will enable considerable simplification of future multi-qubit entanglement quantum circuits that rely on multiple photon sources. Here a ring resonator source with a relatively large radius was used in order to demonstrate wavelength flexibility of a resonant photon source. However, in future work, smaller diameter resonators could be used to increase source brightness and minimize non-degenerate photon noise.

We acknowledge support for this work from the Air Force Research Lab (AFRL). This material is based upon work partially supported by the National Science

[1] J. W. Silverstone, D. Bonneau, K. Ohira, N. Suzuki, H. Yoshida, N. Iizuka, M. Ezaki, C. M. Natarajan, M. G. Tanner, R. H. Hadfield, V. Zwiller, G. D. Marshall, J. G. Rarity, J. L. O'Brien, and M. G. Thompson, Nat. Photonics **8**, 104 (2013).

[2] N. C. Harris, D. Grassani, A. Simbula, M. Pant, M. Galli, T. Baehr-jones, M. Hochberg, D. Englund, D. Bajoni, and C. Galland, Phys. Rev. X **4**, 041047 (2014).

[3] D. Grassani, S. Azzini, M. Liscidini, M. Galli, M. J. Strain, M. Sorel, J. E. Sipe, and D. Bajoni, Optica **2**, 88 (2015).

[4] F. Najafi, J. Mower, N. C. Harris, F. Bellei, A. Dane, C. Lee, X. Hu, P. Kharel, F. Marsili, S. Assefa, K. K. Berggren, and D. Englund, Nat. Commun. **6**, 1 (2015).

[5] C. Schuck, W. H. P. Pernice, and H. X. Tang, Sci. Rep. **3**, 1893 (2013).

[6] S. Azzini, D. Grassani, M. J. Strain, M. Sorel, L. G. Helt, J. E. Sipe, M. Liscidini, M. Galli, and D. Bajoni, **20**, 325 (2012).

[7] R. Wakabayashi, M. Fujiwara, K. Yoshino, Y. Nambu, M. Sasaki, and T. Aoki, Opt. Express **23**, 1104 (2015).

[8] E. Engin, D. Bonneau, C. M. Natarajan, A. S. Clark, M. G. Tanner, R. H. Hadfield, S. N. Dorenbos, V. Zwiller, K. Ohira, N. Suzuki, H. Yoshida, N. Iizuka, M. Ezaki, J. L. O'Brien, and M. G. Thompson, Opt. Express **21**, 27826 (2013).

[9] S. Clemmen, K. P. Huy, W. Bogaerts, R. G. Baets, P. Emplit, and S. Massar, Opt. Express **17**, 16558 (2009).

[10] S. Preble, M. Fanto, C. Tison, Z. Wang, J. Steidle, and P. Alsing, in *SPIE Defense, Secur. Sens.* (Baltimore, MD, 2015).

[11] W. C. Jiang, X. Lu, J. Zhang, O. Painter, and Q. Lin, arXiv:1210.4455v1 (2012).

[12] E. Engin, D. Bonneau, C. M. Natarajan, A. S. Clark, M. G. Tanner, R. H. Hadfield, S. N. Dorenbos, V. Zwiller, K. Ohira, N. Suzuki, H. Yoshida, N. Iizuka, M. Ezaki, J. L. O'Brien, and M. G. Thompson, Opt. Express **21**, 27826 (2013).

[13] D. A. G. Rassani, S. T. A. Zzini, M. A. L. Iscidini, M. A. G. Alli, M. I. J. S. Train, M. A. R. C. S. Orel, J. E. S. Ipe, and D. A. B. Ajoni, Optica **2**, 88 (2015).

[14] Y. Guo, W. Zhang, S. Dong, Y. Huang, and J. Peng, Opt. Lett. **39**, 2526 (2014).

[15] C. Schuck, W. H. P. Pernice, and H. X. Tang, Sci. Rep. **3**, 1893 (2013).

[16] J. W. Silverstone, R. Santagati, D. Bonneau, M. J. Strain, M. Sorel, J. L. O. Brien, and M. G. Thompson, arXiv:1410.8332v4 (2014).

[17] J. W. Silverstone, D. Bonneau, K. Ohira, N. Suzuki, H. Yoshida, N. Iizuka, M. Ezaki, C. M. Natarajan, M. G. Tanner, R. H. Hadfield, V. Zwiller, G. D. Marshall, J. G. Rarity, J. L. O'Brien, and M. G. Thompson, Nat. Photonics **8**, 104 (2013).

[18] Y. Zhang, S. Yang, A. E.-J. Lim, G.-Q. Lo, C. Galland, T. Baehr-Jones, and M. Hochberg, Opt. Express **21**, 1310 (2013).

[19] Z. Vernon and J. E. Sipe, arXiv:1502.05900 (2015).